 \documentclass[aps,prd,preprint,floatfix,nofootinbib,superscriptaddress,showpacs]{revtex4}
 \usepackage{graphicx}
\def\Z0{${\em Z^0\/}$}

\def\r#1 {$^{#1}$}

\newcommand{\gevc} { {\rm GeV/c}}
\newcommand{\gevcc}{ {\rm GeV/c^2}}

\def\gepsfcentered#1{
  \def\testit{#1}
  \def\lbracket{[}
  \ifx\testit\lbracket
    \let\dofilecmd=\gepsfwithopt
  \else
    \let\dofilecmd=\gepsfnoopt
  \fi
  \dofilecmd}

\def\gepsfnoopt#1{
  \begin{center}
  \leavevmode
  \epsffile{#1}
  \end{center}}

\def\gepsfwithopt#1 #2 #3 #4]#5{
  \begin{center}
  \leavevmode
  \gepsfmaxx=0.94\textwidth
  \epsffile[#1 #2 #3 #4]{#5}
  \end{center}}

\newdimen\gepsfmaxx
\def\epsfsize#1#2{
  \ifnum \epsfxsize=0
    \ifnum \epsfysize=0
      \ifnum #1 > \gepsfmaxx
        \gepsfmaxx
      \else
        #1
      \fi
    \else
      \epsfxsize
    \fi
  \else
    \epsfxsize
  \fi
}

\begin{document}

 \bibliographystyle{apsrev}

 \title {Status of the observed and predicted {\boldmath $b\bar{b}$}
         production at the Tevatron}

 \affiliation{Laboratori Nazionali di Frascati, Istituto Nazionale 
              di Fisica Nucleare, Frascati, Italy\\}
 \affiliation{University of Cyprus, 1678 Nicosia, Cyprus\\}

 \author{F.~Happacher} 
 \affiliation{Laboratori Nazionali di Frascati, Istituto Nazionale 
              di Fisica Nucleare, Frascati, Italy\\}
 \author{P.~Giromini} 
 \affiliation{Laboratori Nazionali di Frascati, Istituto Nazionale 
              di Fisica Nucleare, Frascati, Italy\\}
  \author{F.~Ptohos}
 \affiliation{University of Cyprus, 1678 Nicosia, Cyprus\\}
 \begin{abstract}
  We review the experimental status of the $b$-quark production at the 
  Fermilab Tevatron. We compare all available measurements to perturbative
  QCD predictions (NLO and FONLL) and also to the parton-level cross section
  evaluated with parton-shower Monte Carlo generators. We examine both
  the single $b$ cross section and the so called $b\bar{b}$ correlations.
  The review shows that the experimental situation is quite complicated
  because the measurements appear to be inconsistent among themselves.
  In this situation, there is no solid basis to either claim that
  perturbative QCD is challenged by these measurements or, in contrast, 
  that long-standing discrepancies between data and theory have been
  resolved by incrementally improving the  measurements and the theoretical
  prediction. 
 \end{abstract} 
 \pacs{13.85.Lg, 14.65.Fy}
 \preprint{FERMILAB-PUB-05-376-E-T}  
 \maketitle
 \section {Introduction}  \label{sec:ss-intro}
  The bottom quark production at the Fermilab Tevatron has been called one 
  of the few instances in which experimental results appear to challenge the
  ability of perturbative QCD to accurately predict absolute rates in hadronic
  collisions. In general, the data are underestimated by the exact 
  next-to-leading-order (NLO) QCD prediction~\cite{nde,mnr}. The most recent 
  measurement from the Tevatron~\cite{psi-runii} is however in very good
  agreement with an improved QCD calculation (FONLL~\cite{fonll}), and has
  prompted a number of studies~\cite{cacc,mlm,mc} suggesting that the apparent
  discrepancy has been resolved with incremental improvements of the
  measurements and predictions.

  Because of the experimental difficulty inherent to each  result,
  in Sec.~\ref{sec:single-b} we review all measurements of the single $b$
  cross section performed at the Tevatron, and then compare their average
  to the standard and to the improved QCD predictions.
  In Sec.~\ref{sec:show-mc} we review a number of measurements that compare
  the heavy flavor production at the Tevatron to the prediction of 
  parton-shower Monte Carlo generators~\cite{herwig,pythia}.
  Section~\ref{sec:2b} compares cross sections for producing both $b$ and 
  $\bar{b}$ quarks - centrally and above a given transverse momentum cut -
  to theoretical predictions.
  Our conclusions are presented in Sec.~\ref{sec:concl}.
 \section{Single {\boldmath $b$}-quark production cross section at the Tevatron}
 \label{sec:single-b}
 The single $b$-quark cross section is inferred from the measurement
 of the production rate as a function of the transverse momentum, $p_T$,
 of: $B$ hadrons; or some of their decay products (leptons or $\psi$ mesons);
 or jets produced by the hadronization of $b$ quarks. Most of the Tevatron
 measurements correspond to $b$ quarks produced centrally (rapidity 
 $|y^b| \leq 1$) and with $p_T \geq 6 \; \gevc$ 
 (up to $p_T \simeq 100 \; \gevc$).
 The measured cross sections are tabulated as a function of the
 transverse momentum of different objects such as the parent $b$ quark,
 the $B$ hadron, or a $B$-hadron decay prong (a lepton or a $\psi$ meson). 
 This makes the comparison of different measurements quite difficult, and
  usually only a few of them are presented together in review
 articles~\cite{mlm,mc,ri} and  compared to the same theoretical prediction.
 Therefore, we start this review with a consistency check of all available
 data. For that purpose, we use  the value of the observed cross section
 integrated from the $p_T$ threshold of each experiment. We determine the
 ratio $R$ of each measurement to the same theoretical prediction. We then
 evaluate the average $R$ and its dispersion. As standard theoretical 
 prediction, we use the exact NLO calculation~\cite{nde} implemented with a 
 non-perturbative model for the $b$ fragmentation~\footnote{We use a $b$-quark
 mass $m_b=4.75 \; \gevcc$ and renormalization and factorization scales
 $\mu_R=\mu_F=\sqrt{p_T^2 +m_b^2}$.}. The fragmentation model is based on
 the Peterson fragmentation function~\cite{pet} with the $\epsilon$ parameter
 set to 0.006 according to fits to $e^+e^-$ data~\cite{chrin}. When 
 appropriate, $B$ hadrons are decayed using the {\sc qq} Monte Carlo generator
 program~\cite{qq}. The standard prediction is based on old but consistent
 sets of parton distribution functions (PDF) [MRSD$_0$~\cite{mrsd0} and
 MRSA$^{'}$~\cite{mrsa}] since they have been used in most published works. 
 There are 10 measurements of the single $b$ cross section performed by the
 CDF and D${\not\! {\rm O}}$ collaborations at the Tevatron~\footnote{
 We do not include the measurements in Refs.~\cite{cdf1,d01} because they
 are based on a handful of events. The measurement in Ref.~\cite{cdf2} is not
 included because prompt $\psi$ mesons are not separated from those produced
 by $b$-quark decays.}.
 We evaluate the ratio $R$ to the standard theory for the few cases in which
 it is  not provided in the publication. The $b$ cross sections derived from
 the production and decay of $B$ hadrons depend on the value of $f_u$, 
 the fragmentation fraction, and  the branching fractions of the $B$ decay
 available at the time of publication. The value of these parameters has
 changed appreciably over time; we use the same parameters, the value of
 which will be specified in the following, for all measurements and correct
 accordingly the published cross sections. The measurements are based upon
 different $b$-quark signatures:

 The measurement in Ref.~\cite{cdfb1} uses $B$ mesons reconstructed through 
 the decay $B \rightarrow J/\psi \; K $ with $J/\psi \rightarrow \mu^+ \mu^-$.
 The ratio $R= 3.5 \pm 15$\%~\footnote{
 The paper quotes  a discrepancy of $2.9 \pm 15$\% with respect to the
 standard theory that uses the MRST~\cite{mrst-n} set of parton distribution
 functions. This discrepancy is evaluated by fitting the ratio of the data
 to the standard theory as a function of the $B$ $p_T$. This procedure 
 underestimates the ratio of the observed to predicted integrated
 cross sections that is $3.2 \pm 15$\%; this ratio becomes $R= 3.5 \pm 15$\%
 when using the MRSD$_0$ set of structure functions as in the measurement
 described next.}
 for $b$ quarks with $p_T^{\rm min} \geq 9 \; \gevc$ is derived using a
 fragmentation fraction $f_u=0.375$ and a branching fraction of
 $5.88 \times 10^{-5}$~\cite{pdg02}.

 Reference~\cite{cdfb2} is an earlier CDF measurement that uses the same
 decay mode and the same kinematic selection. Using the same fragmentation
 and  branching fractions of the previous measurement, we derive
 $R= 2.9\pm 23$\%~\footnote{
 This ratio is larger than that quoted in the publication (1.9 $\pm$ 15\%)
 and derived  by fitting the ratio of the data to the standard theory as
 a function of the $B$ $p_T$.}.

 Reference~\cite{cdfb3} presents a measurement based on the process 
 $p \bar{p} \rightarrow \mu X$. The contribution of misidentified muons and
 of $c$ quarks is evaluated using Monte Carlo simulations. Using a $b$-quark
 semileptonic branching fraction of 11.2\%, the measurement yields 
 $R= 2.5 \pm 26$\% for $b$ quarks with $p_T^{\rm min} \geq 21 \; \gevc$; for 
 $p_T^{\rm min} \geq 29 \; \gevc$, the ratio is $R= 1.9 \pm 35$\%~\footnote{
 The published result uses the DFLM fits~\cite{dflm}. This old set of parton
 distribution functions is quite similar to the most recent PDF fits and
 yields theoretical cross sections that are
 17\% (for  $p_T^{\rm min} \geq 21 \; \gevc$) and 11\% 
 (for  $p_T^{\rm min} \geq 29 \; \gevc$)
 higher than those obtained using the MRSD$_0$ fits.}.

 Reference~\cite{cdfb4} reports two complementary measurements that use
 the processes $p \bar{p} \rightarrow e\;D^0 X$, with
 $D^0 \rightarrow K^+ \pi^-$, and $p \bar{p} \rightarrow e \; X$; 
 while the first channel is almost background free, the second has large 
 background contributions of misidentified electrons and of electrons from
 $c$-quark decays. These background contributions are determined by studying
  the distribution of
 $p_T^{\rm rel}$, the transverse momentum of the electron with respect
 to the direction of the momentum of all tracks around the electron direction.
 Using a 11.2\% semileptonic branching fraction, 
 the inclusive electron channel yields  $R= 2.4 \pm 27$\% for $b$ quarks with
 $p_T^{\rm min} \geq 15 \; \gevc$. Using a branching fraction of
 $3.14 \times 10^{-3}$, the $e\; D^0$ channel yields $R= 2.1 \pm 34$\%
 for $b$ quarks with  $p_T^{\rm min} \geq 19 \; \gevc$~\footnote{
 The published ratios are based on the use the DFLM fits. We correct for the
 fact that these fits yield a theoretical prediction that is 22\% and 18\%
 larger, respectively, than that based on the MRSD$_0$ fits.}.  
 
 The study in Ref.~\cite{cdfb5} uses the decay $B \rightarrow J/\psi\; X$ with
 $ J/\psi \rightarrow \mu^+ \mu^-$. The $B$ contribution is separated
 from prompt $ J/\psi$ production  by
 studying lifetime distributions. By using the fragmentation fraction
 $f_u=0.375$ and a branching fraction of $6.74 \times 10^{-4}$, the
 measurement yields $R=2.0 \pm$10\% for $b$ quarks with 
 $p_T^{\rm min} \geq 9 \; \gevc$; for $p_T^{\rm min} \geq 14 \; \gevc$,
 $R$ decreases to $1.7 \pm 15$\%.

 Reference~\cite{psi-runii} reports the first  measurement at
 $\sqrt{s}=1.96$ TeV~\footnote{
 All other measurements considered in this review are performed at 
 $\sqrt{s}=1.8$ TeV.}
 through the decay $B \rightarrow J/\psi\; X$ with 
 $ J/\psi \rightarrow \mu^+ \mu^-$. This measurement extends the differential
 cross section to $p_T \simeq 0\; \gevc$, and the data are compared only to
 an improved QCD calculation~\cite{cacc}. We compare to the standard theory
 using the information that the observed cross section is 85\% of that 
 reported in Ref.\cite{cdfb5}, whereas it should have been 10\% 
 larger~\cite{psi-runii}. We derive a ratio  $R=1.5 \pm 9$\% for
 $b$ quarks with $p_T^{\rm min} \geq 9 \; \gevc$;  
 for $p_T^{\rm min} \geq 14 \; \gevc$, $R$ decreases to $1.3 \pm 9$\%.
 
 The study in Ref.~\cite{d0b1} uses the channel $p\bar{p}\rightarrow \mu \; X$.
 The $b$ contribution is separated from  backgrounds due to misidentified
 muons or $c$-quark decays  by looking at the distribution of
 $p_T^{\rm rel}$, the transverse momentum of the muon with respect
 to the direction of a jet with $E_T \geq 8 $ GeV that includes the muon.
 Using a semileptonic branching fraction of 11.2\% and the 
 MRSD$_0$ set of parton distribution functions, the measurement yields
 $R=2.1 \pm 27$\% for $b$ quarks with $p_T^{\rm min} \geq 6 \; \gevc$;
 the ratio is $1.7 \pm 30$\% for $p_T^{\rm min} \geq 12 \; \gevc$.

 Reference~\cite{d0b2} is a repetition of the previous measurement
 that uses slightly different kinematic cuts and an improved simulation of the 
 $b$ hadronization and decay. The publication uses the MRSR2 fits~\cite{mrsr2}.
 We correct for the fact that the theoretical cross sections are 36\% and 18\%
 higher than those obtained using the MRSD$_0$ fits for 
 $p_T^{\rm min} \geq 6 \; \gevc$ and $p_T^{\rm min} \geq 12 \; \gevc$,
 respectively. The measurement yields  $R=2.5 \pm 25$\% for $b$ quarks 
 with $p_T^{\rm min} \geq 6 \; \gevc$ and $R=3.5 \pm 25$\% for $b$ quark with
 $p_T^{\rm min} \geq 12 \; \gevc$.

 Reference~\cite{d0b3} compares the production of central $b$ jets to the
 prediction of the standard theory. The measurement requires the presence
 of a muon within the jets and uses its $p_T^{\rm rel}$ distribution to
 separate the $b$-quark contribution from the background.
 The publication uses the MRSA$^{'}$ fits, and reports $R=2.4 \pm 20$\%
 for $b$ quarks with $p_T^{\rm min} \geq  20\; \gevc $ ; the ratio decreases
 to $R \simeq 2.0 \pm 30$\% for $p_T^{\rm min} \geq 40 \; \gevc $.

 The ratios of the data to the standard theory are summarized in
 Table~\ref{tab:comp}. 
 \begin{table}
 \caption{Ratio $R$ of measured single $b$ cross sections to a prediction
          based on the exact NLO calculation (see text). The cross sections
          are for producing $b$ quarks above a given transverse momentum
          $p_T^{\rm min}$. The ratios in parentheses highlight those cases in
          which data and theory appear to have different transverse momentum 
          distributions and are not used in deriving $<R>$. Each measurement
          covers $b$-quark momenta as large as $4-5$ times the $p_T^{\rm min}$
          threshold. The measurement in the seventh row also covers small
           transverse momenta down to $p_T \simeq 0\; \gevc$.} 
\begin{center}
\begin{ruledtabular}
\begin{tabular}{lccccccc}
 channel& (experiment) & \multicolumn{6}{c}{$R$ for   $p_T^{\rm min}$ ($\gevc$) $\geq$ } \\
       &&  6& $8-10$  &  $12-15$ & $19-21$   & $\simeq 29$  & $\simeq 40 $ \\
 $J/\psi K^+$& (CDF~\cite{cdfb1})  &   &  $3.5\pm 15$\%   &  (3)               &              &   &    \\
  $J/\psi K^+$& (CDF~\cite{cdfb2}) &   &  $2.9\pm 23$\%   &  (1.9)              &              &   &     \\
 $\mu \;X$& (CDF~\cite{cdfb3})     &   &                 &      &$2.5\pm 26$\%    &  (1.9)         &     \\
  $e \;X$& (CDF~\cite{cdfb4})      &   &                 &$2.4\pm 27$\%    &         &   &        \\
 $e D^0$& (CDF~\cite{cdfb4})       &   &                 &  & $2.1\pm 34$\%      &   &        \\
 $J/\psi \; X $& (CDF~\cite{cdfb5}) &  & $2.0\pm 10$\% & (1.7) &     &   &   \\
  $J/\psi \; X $& (CDF~\cite{psi-runii}) &  & $1.5\pm 9$\% & (1.3) &     &   &   \\
 $\mu \; X$& (D${\not\!{\rm O}}$~\cite{d0b1}) & $2.1\pm 27$\% & &(1.7)  & & &  \\
 $\mu \; X$& (D${\not\!{\rm O}}~$\cite{d0b2}) & $2.5\pm 25$\% & & (3.5)  & & &  \\
 $b$ jets ($\mu$)& (D${\not\!{\rm O}}$~\cite{d0b3})   & & & &  $2.4\pm 20$\% &  & (2.0) \\
\end{tabular}
 \end{ruledtabular}
 \end{center}
 \label{tab:comp}
 \end{table}
 Using the measurements listed in Table~\ref{tab:comp}, we derive an average
 ratio of the data to the standard theory that is $<R>=2.39$; 
 the RMS deviation of the 10 measurements in Table~\ref{tab:comp} is 0.54
 that in turn yields a 0.17 error on $<R>$.
 Before comparing the data to the improved QCD calculation, the summary
 of the experimental situation in Table~\ref{tab:comp} prompts a few
 additional remarks.

 The 0.54 RMS deviation is much larger than the measurement uncertainties
 (these uncertainties are dominated by systematic errors that are generally 
 quoted as conservative estimates). When not using the four measurements 
 based on detection of $J/\psi$ mesons, the average ratio becomes $<R>=2.33$
 with a 0.19 RMS deviation that, as expected, is smaller than the measurement
 uncertainties. The remaining measurements based on detection of $J/\psi$
 mesons (first, second, sixth, and seventh line in Table~\ref{tab:comp})
 yield  $<R>=2.5$ with a RMS deviation of 0.9. These four measurements are
 experimentally the cleanest and easiest to perform, and have the smallest 
 systematic errors; however, they appear to be inconsistent among themselves.
 For the latter reason, it does not seem judicious to use only these four
 measurements  as benchmarks of theoretical progresses~\cite{cacc,cana}.
 Additional data by CDF and D${\not\!{\rm O}}$ are certainly needed to
 clarify this situation.

 In most, but not all, measurements the shape of the observed transverse
 momentum distribution is different from that of the standard theory
 (see values in parentheses in Table~\ref{tab:comp}). In general, data and
 theory tend to agree better with increasing $p_T$; in one case (ninth line
 of Table~\ref{tab:comp}) they disagree more. It could be a real effect, but
 it remains open the possibility that some measurements do not model correctly
 the background contribution as a function of the $b$-quark transverse
 momentum.

 As noted in Refs.~\cite{cacc,mlm}, the measurement with $b$ jets, listed in
 the last row of Table~\ref{tab:comp}, depends little on the modeling of the
 $b$-quark fragmentation.  This measurement yields $R=2.4 \pm 20$\%, whereas
 $<R>=2.39 $, and does not provide, in contrast to what claimed
 in Refs.~\cite{mlm,mc}, any evidence that the $b$ fragmentation function
 is a major cause of discrepancy between data and standard theory~\footnote{
 This result is confirmed by a recent measurement~\cite{monica} of the single
 $b$ cross that uses central jets with transverse energy $E_T \geq$ 40 GeV;
 $b$ jets are selected identifying the presence of displaced secondary 
 vertices. This study finds that the ratio of the observed cross section to 
 that predicted by the {\sc pythia} Monte Carlo generator is $1.2 \pm 20$\%; 
 this implies that the ratio of these data to the standard NLO prediction is
 approximately 2.3 (see the discussion in the next section).}.

 The NLO prediction depends strongly on the choice of the factorization and
 normalization scales; by changing the scales by a factor of two the 
 prediction changes by approximately 40\%~\cite{nde,mnr,ri}. At perturbation
 level, the large scale dependence of the NLO prediction is generally taken 
 as a symptom of large higher-order corrections~\footnote{
 It is well known that there are new partonic processes that appear first at
 NLO, such as gluons branching into $b$ and $\bar{b}$ quarks (gluon splitting)
 in the final or initial state of the hard scattering.}.
 In addition, there are logarithmic corrections that are present at all
 orders of perturbation theory~\cite{pe1,pe2,pe3,pe4}. The resummation of the
 logarithms of $(p_T/m_b)$ with next-to-leading logarithmic accuracy (NLL), and
 the matching with the fixed-order NLO calculation (FONLL) for massive quark 
 has been performed in Ref.~\cite{fonll}. A calculation with the same level
 of accuracy, available for the production of $b$ quarks at $e^+\;e^-$
 colliders~\cite{f1},  has been used to extract non-perturbative
 fragmentation functions from LEP and SLC data~\cite{f2}. These new 
 fragmentation functions have been consistently~\footnote{
 As correctly noted in Ref.~\cite{cana}, the Peterson form of the fragmentation
 function used in the standard NLO calculation has been tuned in conjunction
 with a parton-level cross section evaluated with the leading-log (LL) 
 approximation of parton-shower Monte Carlo programs and should not be
 convoluted with a NLO prediction.}
 convoluted with the FONLL calculation to predict the $B$ cross section at 
 the Tevatron~\cite{cana}. The inclusion of NLL logarithms has a modest 
 effect in the $p_T$ range considered in this review; the new fragmentation
 functions are harder than the Peterson fragmentation function and explain 
 most of the 30\% increase of the FONLL prediction with respect to the
 standard theory~\cite{cana}. In the $p_T$ range considered in this study,
 the latest PDF fits, that include HERA data~\cite{mrst2001,cteq} and a more 
 accurate value of $\alpha_s$ at the $Z$ mass, increase by 20\% the predicted
 $b$-quark cross section with respect to the PDF fits used for the comparisons
 in Table~\ref{tab:comp}~\cite{mlm}. By also using $f_u=0.39$ in place of
 0.375, the final FONLL prediction is approximately 60\% higher than the
 standard NLO prediction. In conclusion, the ratio of the average single 
 $b$ cross section measured at the Tevatron with respect to the FONLL
 prediction is approximately 1.5. The uncertainty of the FONLL prediction is
 estimated to be approximately 40\%~\cite{cacc}~\footnote{
 The uncertainty is estimated by changing the normalization and factorization
 scales by a factor of two ($\pm 35$\%) and $m_b$ by $\pm 0.25 \; \gevcc$
 ($\pm 16$\%).}. 
 Therefore, the average single $b$ cross section measured at the Tevatron is
 within the range of values predicted by the FONLL calculation.

 Exact NLO predictions do not easily allow the full simulation of events
 produced at the Tevatron. Therefore, studies that involve $b$-quark
 production such as top quark studies or searches for new physics, make use
 of parton-shower Monte Carlo programs~\cite{herwig,pythia}. Parton-level
 cross sections, evaluated by these generators using the leading-log (LL)
 approximation, also have large uncertainties, comparable
 to that of the NLO or FONLL prediction:  gluon splitting to heavy
 quarks in the final state has a 30\% uncertainty~\cite{seymour} whereas
 gluon splitting in the initial state (flavor excitation diagrams) depends on
 the PDF choice and can vary by as much as $\pm 40$\% when using a wide range
 of structure functions in the PDF library~\cite{pdf}. Since, as correctly 
 noted in Ref.~\cite{mlm}, studies searching for new physics cannot depend
 on the prediction of a QCD calculation, a significant effort has been put in
 calibrating the parton-level cross section predicted by parton-shower Monte
 Carlo programs by using jet data~\cite{ajets,super,xsec}. Buried in top 
 quark studies or dubious hints of new physics, the significance of
 this calibration has been overlooked. Therefore, we review it in detail 
 in the next section.
 \section{Comparisons  with the herwig and  pythia predictions}
 \label{sec:show-mc}
 It was first reported in Ref.~\cite{field} that parton-shower Monte Carlos,
 such as the {\sc pythia} and {\sc herwig} generators, predict a parton-level
 single $b$ cross section that approximately matches the Tevatron measurements
 for $b$ quarks with $p_T \geq 6 \; \gevc$ and  $|y|\leq 1$. 
 The parton-level cross section estimated with LL generators is approximately
 a factor of two larger than the exact NLO prediction because the contribution
 of terms of order higher than $\alpha_s^2$ is a factor of two larger than the
 contribution of $\alpha_s^3$ terms estimated with the exact NLO
 calculation~\cite{mnr}.
 
 Leading-order (LO) and higher-than-LO terms are sources of $b$ and $\bar{b}$
 quarks with quite different topological structure. The production of events
 with both a $b$ and  $\bar{b}$ quark with $p_T \geq 6 \; \gevc$ and 
 $|y| \leq 1$ is dominated by  LO diagrams and the parton-level cross
 sections predicted by the exact NLO calculation is comparable to that
 predicted by LL Monte Carlo generators. At the time, both LL and NLO
 predictions underestimated by a factor of two the available 
 measurements~\cite{d0b2,derw,2mucdf}~\footnote{
 These measurement identify $b$ quarks through their semileptonic decays.}.
 Therefore, the fact that LL generators model correctly the single
 $b$ cross section was considered merely accidental, and the source of
 the discrepancy between data and NLO prediction was searched in
 non-perturbative fragmentation effects that enhance equally LO and NLO terms.

 In Refs.~\cite{ajets,xsec}, the heavy flavor cross section evaluated with
 the {\sc herwig} Monte Carlo generator has been tuned by using jet data
 collected by the CDF experiment at the Tevatron. This study uses four
 samples of data consisting of events with two or more jets, one of which
 is central and has transverse energy $E_T$ larger than 20, 50, 70, and
 100 GeV, respectively, and a data sample, richer in heavy flavor, collected
 requiring two or more central jets with  $E_T \geq 15$ GeV, one of which
 contains a lepton with $p_T \geq 8 \; \gevc$ from heavy-flavor decays. 
 Jets containing heavy flavor are identified by finding displaced secondary
 vertices produced by the decay of $b$ and $c$ hadrons inside a jet;
 an additional algorithm uses track impact parameters to select jets with
 a small probability of originating from the primary event vertex. 
 In the data, the $b$- and $c$-quark contributions are separated because both
 algorithms have the same tagging efficiency for $b$ jets, whereas for $c$ 
 jets the efficiency of the second algorithm is approximately 2.5 times larger
 than that of the first algorithm. The tagging rates in the data are compared
 to those of samples simulated using the {\sc herwig} Monte Carlo
 program~\footnote{
 The study uses option 1500 of version 5.6 with the MRS(G) set of parton
 distribution functions~\cite{mrsa}.}.
 The study compares momentum distributions of leptons or of the system of 
 tracks forming a secondary vertex (decay products of the $B$ hadron inside
 the jet) in the data and simulation. This comparison shows that the 
 hadronization of heavy quarks at the Tevatron is modeled correctly by 
 {\sc herwig} tuned with $e^+\;e^-$ data. Therefore, one is allowed to tune
 the parton-level cross section predicted by the Monte Carlo generator to
 match the heavy-flavor content - or the tagging rate - of the data. The
 contribution of LO and higher-order terms can be separated because the 90\% 
 of the LO contribution consists of events which contain two jets with heavy
 flavor inside the kinematic cuts. In contrast, only 10\% of the events due 
 to higher-order terms contains two jets with heavy flavor in the detector
 acceptance. The higher-order contributions due to gluons splitting into
 heavy quarks in the initial and final state are disentangled by studying
 the $\delta R=\sqrt{(\delta \phi)^2+(\delta \eta)^2}$ distribution between
 two jets with heavy flavor (gluon splitting in the final state clusters 
 at small values of $\delta R$).    

 References~\cite{ajets,xsec} show that the data can be modeled by tuning
 the parton-level cross section predicted by the {\sc herwig} generator
 within the theoretical and experimental uncertainties~\footnote{
 The gluon splitting in the final state predicted by {\sc herwig} has to be
 increased by ($40\pm 20$)\%. Before tuning the simulation, the size of
 gluon splitting in the final state predicted by {\sc herwig} is 1/2 of
 that in the initial state.}.
 In the tuned LL generator, the contribution of higher-order terms to the 
 single $b$ cross section is approximately four times larger than the LO 
 contribution. In contrast, for the same kinematics, the exact NLO
 calculation with standard scales returns $\alpha_s^3$ contributions that
 are only a factor of two larger than the $\alpha_s^2$ contribution. 
 As discussed in the next section, the study of $b\bar{b}$ correlations can 
 be used to assess the correct ratio of higher-than-LO to LO contributions.
 \section{Measurement of the {\boldmath $b \bar{b}$} correlations}
 \label{sec:2b}
 The cross section for producing both $b$ and $\bar{b}$
 quarks centrally and above a given $p_T$ threshold, $\sigma_{b\bar{b}}$
 or $b\bar{b}$ correlation, is dominated by LO terms, and
 the LL and NLO predictions are quite close~\footnote
 {For example, in Ref~\cite{ajets} the tuned LL generator predicts a modest
 contribution of higher-than-LO order terms to $\sigma_{b\bar{b}}$
 ($\simeq 30$\%); for the same kinematics, the exact NLO calculation predicts
 a $\simeq 15$\% contribution of higher-than-LO order terms. In this case, 
 the LL and NLO predictions of $\sigma_{b\bar{b}}$ are within 20\%.}.
 In addition, the exact NLO prediction of $\sigma_{b\bar{b}}$ depends little 
 on the choice of the normalization and factorization scales as well as on the
 $b$-quark mass~\footnote{
 The prediction changes by no more than 15\% by changing the scales by a 
 factor of two and $m_b$ by $\pm$ 0.25 $\gevcc$~\cite{ajets}.}
 and appears to be a robust prediction of perturbative QCD.

 Therefore, it is important to determine precisely the value of $R_{2b}$,
 the ratio of $\sigma_{b\bar{b}}$ measured at the Tevatron to the exact NLO
 prediction (or to the LL prediction that is very close). A ratio
 $R_{2b} \simeq 1$ would imply that the parton-level cross section predicted
 by LL Monte Carlo generators is correct and  that the contribution of
 higher-than-LO terms has to be a factor of two larger than in the present
 NLO or FONLL prediction. If the ratio $R_{2b}$ is much larger than one,
 then the agreement between the observed single $b$ cross section and the
 prediction of LL Monte Carlo generators is fortuitous. Since the
 NLO prediction of $\sigma_{b\bar{b}}$ is robust, agreement with the data
 may be found by using harder fragmentation functions as in the FONLL 
 calculation. Unfortunately, the status of the $\sigma_{b\bar{b}}$
 measurements at the Tevatron is quite disconcerting.

 The study in Ref.~\cite{ajets}(CDF) uses two central jets with $E_T \geq 15$
 GeV, each containing a secondary vertex due to $b$- or $\bar{b}$-quark decays.
 The LL prediction, tuned to fit the data,  yields $R_{2b}=1.2$~$^{15}$ with
 a 25\% uncertainty mostly due to the efficiency for finding a secondary
 vertex in a jet.

 A recent measurement~\cite{shears} (CDF) supports the conclusion of
 Ref.~\cite{ajets}. The study in Ref.~\cite{shears} uses events containing
 two central jets with $E_T\geq 30$ and 20 GeV, respectively; pairs of $b$
 jets are also identified by requiring the presence of displaced secondary
 vertices. This study finds that the ratio of  $\sigma_{b\bar{b}}$ to the LL
 {\sc pythia} prediction is $0.9\pm 31$\%, while the ratio of 
 $\sigma_{b\bar{b}}$ to the NLO prediction~\footnote{
 In this case the NLO prediction has been evaluated using the 
 {\sc MC@NLO} Monte Carlo generator~\cite{webber}.}
 is  $R_{2b}=1.0 \pm 32$\%.
   
 In contrast, discrepancies between data and the NLO prediction of 
 $\sigma_{b\bar{b}}$ are observed when identifying $b$ quarks through their
 semileptonic decay into muons.

 The study in Ref.~\cite{derw}(CDF) uses events with a muon recoiling against
 a jet that contains tracks with large impact parameter ($b$ jet). Using the 
 average branching fraction BR = 10.3 for $b \rightarrow \mu\; X$ decays
 and 10.2 for $b \rightarrow c \;X \rightarrow \mu \; Y$  sequential 
 decays~\cite{pdgold}, the ratio of $\sigma_{b\bar{b}}$ to the exact NLO 
 prediction  is measured to be $R_{2b}=1.5 \pm 10$\% for $b$ and $\bar{b}$
 quarks produced centrally and with $p_T^{\rm min} \geq 12\; \gevc$.

 Reference~\cite{2mucdf} (CDF) reports a measurement that uses two centrally
 produced muons. By using the square of the semileptonic branching fraction
 quoted above, the study yields $R_{2b}=3.0 \pm 20$\% for central $b$ and 
 $\bar{b}$ quarks with $p_T^{\rm min}\geq 6\; \gevc$.

 Reference~\cite{d0b2} (D${\not\!{\rm O}})$ reports an analogous measurement
 that also uses two centrally produced muons. The square of the semileptonic
 branching fraction is evaluated with the {\sc isajet} generator~\cite{isajet}
 implemented with the {\sc qq} decay table and is consistent with the value
 used by CDF. The study yields a ratio $R_{2b}=2.3 \pm 33$\% for central $b$
 and $\bar{b}$ quarks with $p_T^{\rm min} \geq 7\; \gevc$.

 These five measurements, listed in Table~\ref{tab:comp_1}, 
 yield $<R_{2b}>=1.8$ with a 0.8 RMS deviation. Such a large RMS deviation
 indicates that the experimental results are inconsistent among themselves. 
 Additional measurements are certainly needed to clarify the experimental
 situation.
 \begin{table}
 \caption{Ratio $R_{2b}$ of $\sigma_{b\bar{b}}$, the observed cross section
	  for producing both $b$ and $\bar{b}$ quarks, centrally and above a 
          given $p_T^{\rm min}$ threshold, to the exact NLO prediction (see 
          text). Each measurement covers $b$-quark momenta as large as $4-5$
          times the $p_T^{\rm min}$ threshold. Jets produced by $b$ and 
          $\bar{b}$ quarks are identified by the presence of displaced 
          secondary vertex or tracks with a large impact parameter. Muons from
          $b$ and $\bar{b}$ decays are separated from the background by 
 	  studying impact parameter~\cite{2mucdf} or 
          $p_T^{\rm rel}$~\cite{d0b2} distributions.}
\begin{center}
\begin{ruledtabular}
\begin{tabular}{lccccc}
 channel& (experiment) & \multicolumn{4}{c}{$R_{2b}$ for $p_T^{\rm min}$ ($\gevc$) $\geq$ } \\
    &  &  $6-7$  &  $10$ & $15$   & $\simeq 20$   \\
 $b +\bar{b}$ jets& (CDF~\cite{ajets})  &   & &   $1.2 \pm 25$\%   &       \\
  $b +\bar{b}$ jets& (CDF~\cite{shears})  &   & & &  $1.0 \pm 32$\%          \\
 $\mu + b$ jet & (CDF~\cite{derw})  &   &   $1.5\pm 10$\%    &           &     \\
  $\mu^+ + \mu^-$ & (CDF~\cite{2mucdf}) &$3.0\pm 20$\%    &         &   &        \\
  $\mu^+ + \mu^-$ & (D${\not\!{\rm O}}$~\cite{d0b2}) &$2.3\pm 33$\%    &         &   &        \\
\end{tabular}
 \end{ruledtabular}
 \end{center}
 \label{tab:comp_1}
 \end{table}
 However, it is quite obvious that the present discrepancies are reduced if
 the rate of observed  semileptonic decays is approximately 50\% higher than
 what is expected because:  
 (a) lepton identification efficiencies are underestimated by approximately
 50\% or (b) additional objects with a 100\% semileptonic branching ratio and
 a cross section of the order of 1/10 of the $b$ cross section are 
 produced~\cite{ajets}.  Reference~\cite{ajets} has investigated these
 hypotheses by comparing the rate of observed and predicted leptons 
 from $b$-quark decays in jets that recoil against a generic jet or a jet 
 that also contains a lepton (the jets are central with $E_T \geq 15$ GeV).
 This study finds that in the second case the rate of jets containing a 
 lepton from presumed $b$ decays is 50\% higher than in the first case.
 The magnitude of the effect is consistent with hypothesis (b). In light
 of this observation, it is worth to go back to Table~\ref{tab:comp}.
 If there was a reason to disregard the first two measurements using the 
 $B \rightarrow J/\psi K$ channel, one would see a completely different
 picture. Six measurements identify $b$ quarks through their semileptonic
 decay and yield $<R>=2.33 \pm 0.06$; the measurements with inclusive 
 $J/\psi$ mesons (sixth and seventh line) do not use $b$ semileptonic decays
 and yield $<R>=1.7 \pm 0.1$; this conjecture very well highlights the need 
 for additional cross-checks of the measurements based on
 $B \rightarrow J/\psi K$ decays and on the inclusive $J/\psi$ production.
 \section{Conclusions} \label{sec:concl}
 We review all measurements of the single $b$ cross section performed at the
 Tevatron and compare them to an exact NLO perturbative QCD prediction, that
 uses pre-HERA sets of parton distribution functions and the Peterson
 fragmentation function, in order to test their consistency.
 We also compare the data to an improved QCD calculation (FONLL) and to
 the prediction of LL Monte Carlo generators. The average ratio of the data to
 the NLO prediction is $<R>=2.39$ with a 0.54 RMS deviation. The RMS deviation
 is much larger than the quoted measurement uncertainties, and indicates that
 experimental results are inconsistent among themselves. With this caveat,
 the average of the data  is found to be in agreement with the parton-level
 cross section evaluated with parton-shower Monte Carlo generators and is
 within the range of uncertainty of the FONLL prediction that in turn is 60\%
 higher than the NLO prediction. The increase of the FONLL prediction with
 respect to the  NLO calculation is mostly due to PDF improvements and the 
 usage of a harder fragmentation function, whereas the parton-level cross
 section are the same in both predictions. On the contrary, the contribution
 of higher-than-LO terms returned by LL Monte Carlo generators fitted to the
 data is approximately a factor of two larger than that in the FONLL or NLO
 calculations. The measurement of $\sigma_{b\bar{b}}$, the cross section for
 producing both $b$ and $\bar{b}$ quarks centrally and above the same $p_T$ 
 threshold, has a decisive role in assessing the correct parton-level cross
 section. In fact, the higher-than-LO contribution to $\sigma_{b\bar{b}}$ is
 quite modest in all theoretical approaches. Because of the use of harder
 fragmentation functions, the FONLL calculation yields a prediction
 appreciably larger than that of LL generators or NLO generators convoluted
 with the Peterson fragmentation function. Unfortunately, the experimental
 situation is quite disconcerting and only raises additional questions. The
 average ratio of the $\sigma_{b\bar{b}}$ measurements to the NLO prediction
 is $<R_{2b}>=1.8$ with a 0.8 RMS deviation, and suggests that these 
 measurements are also inconsistent among themselves. The  $<R_{2b}>$ value
 supports the FONLL approach. However, the level of agreement between data
 and theory appears to be a function of the number of semileptonic
 decays used to identify $b$ quarks. In this situation, it cannot be excluded
 that the $b$ parton-level cross section is correctly described by  LL
 Monte Carlo generators, and that measurements using $b$ semileptonic decays
 are affected by new physics.   


 \end{document}